\documentclass[12pt]{article}
\usepackage{amsmath,amsfonts,amssymb,a4,epsfig}

\newcommand{\R}{{\mathbb{R}}}
\newcommand{\C}{{\mathbb{C}}}
\newcommand{\Z}{{\mathbb{Z}}}

\def\pa{\partial}
\def\ra{\rightarrow}
\def\preuve{\begin{proof}} 
\def\ga{\alpha}

\def\gd{\delta}

\def\go{\omega}

\def\san{San V{\~u} Ng{\d o}c}

\newtheorem{defi}{Definition}[section]

\newtheorem{lemm}{Lemma}[section]

\newtheorem{theo}{Theorem}[section]

\newenvironment{demo}{\noindent {\it Proof.--}
      \begin{quotation}\noindent}{\end{quotation}\hfill$\square $}

\begin{document}

\title{An extension of the Duistermaat-Singer Theorem to
the semi-classical Weyl algebra}
\author{Yves Colin de Verdi\`ere\footnote{Institut Fourier,
 Unit{\'e} mixte
 de recherche CNRS-UJF 5582,
 BP 74, 38402-Saint Martin d'H\`eres Cedex (France);
yves.colin-de-verdiere@ujf-grenoble.fr}}


\maketitle

\begin{abstract}
Motivated by many  recent works (by L. Charles, V. Guillemin, T. Paul,
J. Sj\"ostrand,  A. Uribe, \san,
 S. Zelditch
and others)  on the semi-classical Birkhoff
normal forms, we investigate the structure of the group
of  automorphisms  of the graded semi-classical Weyl algebra.
The answer is quite similar to the Theorem of Duistermaat and Singer
for the usual algebra of pseudo-differential operators where 
 all automorphisms 
 are given by conjugation by an elliptic Fourier Integral
Operator (a FIO).
 Here what replaces general non-linear symplectic diffeomeorhisms
is just linear complex symplectic maps, because everything is localized at
a single point\footnote{Thanks to Louis Boutet de Monvel for his comments
and suggestions, in  particular the proofs of Lemmas
\ref{lemm:derivation} and   \ref{lemm:inner}.
As he said, the main result is not surprising!}.

\end{abstract}
\section{The result}
Let $W=W_0\oplus W_1 \oplus \cdots $ be the 
semi-classical graded  Weyl algebra (see Section \ref{sec:weyl} for a 
definition)
 on $\R^{2d}$. Let us define
$X_j:=W_j\oplus W_{j+1}\oplus \cdots $.
We want to prove the:
\begin{theo}\label{theo:main}
There exists an exact sequence of groups 
\[ 0 \ra {\cal I} \ra_1 {\rm Aut}(W) \ra_2 {\rm Sympl}_\C(2d)\ra 0 \]
where 
\begin{itemize}
\item ${\rm Sympl}_\C(2d)$ is the group of linear symplectic
  transformations
of $\C^{2d}=\R^{2d}\otimes \C $
\item ${\rm Aut}(W) $ is the group of automorphisms $\Phi $ of the
 semi-classical graded\footnote{``graded'' means that
$\Phi( W_n)\subset W_n\oplus W_{n+1} \oplus \cdots $}
  Weyl algebra preserving $\hbar$
\item 
${\cal I}$ is the group  of ``inner'' automorphisms $\Phi_S$
of the form $\Phi_S={\rm exp}(i{\rm ad}S/\hbar )$, i.e.
$\Phi_S (w)={\rm exp}(iS/\hbar )\star w \star {\rm exp}(-iS/\hbar )$ 
as a formal power series, with
$S\in X_3$
\item The arrow $\ra _2 $ is just given from the action of the
  automorphism
$\Phi $ 
on $W_1=\left(\R^{2d}\right)^\star \otimes \C $ modulo $X_2$
\end{itemize}
\end{theo}
The proof follows \cite{D-S} and also the semi-classical version
of it by H. Christianson \cite{Ch}. 
This result is implicitly stated in Fedosov's book \cite{Fed} in   Chapter
5, but it could be nice to have an direct proof in a simpler
context.
The result is a consequence of Lemmas \ref{lemm:surj},
\ref{lemm:derivation} 
and \ref{lemm:inner}.

\section{The Weyl algebra }\label{sec:weyl}
The elements of the ``Weyl algebra'' 
are the  formal power series in $\hbar $ and $(x,\xi) $
\[ W=\oplus_{n=0}^\infty W_n \] 
where $W_n $ is the space of complex valued homogeneous polynomials
in $z=(x,\xi) $ and $\hbar $ of total degree $n$
where   the degree of $ \hbar^{j}z^\ga $
is $2j +|\ga |$.
The Moyal $\star-$product
\[ a\star b:=\sum _{j=0}^{\infty }
 \frac{1}{j! }\left( \frac{\hbar}{2i}
\right) ^j a \left( \sum _{p=1}^d \stackrel{\leftarrow}{\pa} _{\xi_p }
\vec{\pa} _{x_p}
- \stackrel{\leftarrow}{ \pa}_{x_p} \vec{\pa} _{\xi_p} \right)^j b= 
  ab +\frac{\hbar}{2i}\{ a,b \}+ \cdots ~\] 
(where $\{ a,b \} $ is the Poisson bracket of $a$ and $b$)
 gives to $W$
the structure of  a graded algebra:
we have $W_m\star W_n \subset W_{m+n} $ and   hence, for the brackets,  
  $\frac{i}{\hbar }[W_m,W_n]^\star \subset W_{m+n-2}$.

The previous grading of $W$ 
 is obtained by looking at the action of $W$  on the
(graded) vector space ${\cal S}$ of {\it symplectic spinors} (see \cite{Gu}):
if $F\equiv \sum_{j=0}^\infty \hbar ^j F_j (X) $ with $F_j\in {\cal
  S}(\R)$,
we define
$f_\hbar (x)=\hbar^{-d/2}F(x/\hbar)  $ whose micro-support
is the origin. $W $ acts on ${\cal S}$
in a graded way as differential operators of infinite degree:
if $w\in W$, $w.f={\rm OP}_\hbar (w)(f) $.

\section{A remark}

We  assumed in Theorem \ref{theo:main} that $\hbar $
 is fixed by the automorphism.
If not, the symplectic group has to be replaced by the homogeneous
symplectic group: the group  of the linear automorphisms $M$
of $(\R^{2d},\go)$ which satisfies $M^\star \go =c\go $.
We then have to take into account a multiplication of $\hbar $
by $c$. For $c=-1$, it is a semi-classical version of the
transmission property of Louis Boutet de Monvel. 

\section{ Surjectivity of the arrow $\ra _2$}

\begin{lemm}\label{lemm:surj}
The arrow  $\ra _2$ is surjective.
\end{lemm}
\begin{demo}
Let us give $\chi \in {\rm Sympl}_\C(2d)$. The map
$a \ra a\circ \chi $ is an automorphism of the Weyl algebra:
the Moyal formula is given only in terms of the Poisson bracket.
\end{demo}

\section{The principal symbols}

Let $\Phi $ be an automorphism of $W$.
Then $\Phi $ induces a linear automorphism $\Phi_n$ of 
$W_n $: if $w=w_n +r$ with $w_n \in W_n$ and $ r\in X_{n+1}$
and $\Phi(w)=w'_n +r'$ with  $w'_n \in W_n$ and $ r'\in X_{n+1}$, 
$\Phi_n (w_n):=w'_n$ is independent of $r$. 
The polynomial  $w_n$ is the {\it principal symbol} of $w\in X_n$.
We have 
$\Phi_{m+n}(w_m\star  w_n)=\Phi_m(w_m)\star\Phi_n(w_n) $.
Hence $\Phi_n$ is determined by $\Phi_1$ because
the algebra $W$ is generated by $W_1$ and $\hbar$.
The linear map  $\Phi_1$ is an  automorphism of the complexified 
dual of $\R^{2d}$.
Let us show that it preserves the {\it Poisson bracket}
and hence is  the adjoint  of a linear symplectic
mapping of $\C^{2d}$. We have: 
\[ \Phi ([w,w']^\star)=[\Phi (w),\Phi(w')]^\star~.\]
By looking at principal symbols, for $w,w'\in X_1$,
we get
\[  
\{  \Phi _1(w),\Phi _1(w') \}= \{ w,w' \}  ~.\]

\section{Inner automorphisms}\label{sec:inner}
The kernel of $\ra_2$ is the group of automorphisms $\Phi$
which satisfy $\Phi_n ={\rm Id}$ for all $n$, i.e.  for any  $w_n \in W_n$
\[ \Phi (w_n)= w_n~ {\rm mod}~ X_{n+1} ~.\]
 The following fact is certainly well known:
\begin{lemm}\label{lemm:derivation} If $\Phi \in \ker (\ra_2)$,  
$\Phi = {\rm exp}(D)$ where $D:W_n \ra X_{n+1}$
is  a derivation of $W$.
\end{lemm}
{\it Proof [following a suggestion of Louis Boutet de Monvel]--}
 We  define $\Phi^s$ for $s\in\Z$.
 Let  $\Phi^s _{p,n} :W_n \ra W_{n+p}$
be the degree $(n+p)$ component of $ ( \Phi^s)_n: W_n \ra X_{n}$.
Then $\Phi^s_{p,n}$ is polynomial w.r. to $s$. This allows to  extend
$\Phi^s$ to $s\in \R $ as a 1-parameter group
of automorphisms.
We put  $D=\frac{d}{ds}\left(\Phi^s\right)_{|s=0}$. We have
$\Phi^s ={\rm exp}(sD)$.
We deduce that $D$ is a derivation.\hfill $\square$

We need to show the: 
\begin{lemm}\label{lemm:inner}
Every derivation $D$  of $W$ sending $W_1$ into $X_2$ is an
inner derivation of the form 
\[ Dw=\frac{i}{\hbar}[S,w] \]
with $S \in X_3$.
\end{lemm}
{\it Proof [following a suggestion of Louis Boutet de Monvel]--}
 
Let $(\zeta_k )$ the  basis of $W_1$ dual
to the canonical basis $(z_k)$ for the star bracket, i.e. satisfying
$[\zeta_k, z_l]=\frac{\hbar}{i}\gd _{k,l}$.
We have $[\zeta_k ,w]=\frac{\hbar}{i}\frac{\pa w}{\pa z_k}$. 
Put $y_k=D\zeta_k \in X_2$.
As the brackets $[\zeta_k,\zeta_l ]$ are constants,
we have:
$[D\zeta _k, \zeta_l]+[\zeta _k,D \zeta_l] =0$, or
$\pa y_k/\pa z_l=\pa y_l/\pa z_k$.
There exists an unique $S$ vanishing at $0$ so that:
\[ [S, \zeta_k ]=-\frac{\pa S}{\pa z_k}=\frac{\hbar}{i}y_k ~.\]
Hence $D=(i/\hbar ) [S,.] $.
Because $y_k\in X_2$, $S$ is in $X_3$.\hfill $\square$

\section{An homomorphism from 
the group $G$ of elliptic  FIO's
whose associated canonical transformation
fixes the origin  into  ${\rm Aut}(W)$}

 Each  $w\in W$ is  the Taylor expansion
of a Weyl symbol $a\equiv \sum_{j=0}^\infty  \hbar ^j a_j(x,\xi) $
of a pseudo-differential operator $\hat{a}$.
Let us give an elliptic  Fourier Integral Operator $U$ associated to a
canonical transformation $\chi $ fixing the origin.
The map $\hat{a} \ra U^{-1}\hat{a} U$ induces a map $F$
from $S^0 $ into $S^0 $ which is an automorphism of algebra 
(for the Moyal product).
\begin{lemm}
The Taylor expansion of $F(a)$ only depends on the Taylor expansion
of $a$.
\end{lemm}
This is clear from the explicit computation  and the  stationary 
phase expansions.

As a consequence, $F$ induces an automorphism $F_0$  of the 
Weyl algebra graded by powers of $\hbar$.
\begin{lemm}
$F_0 $ is an automorphism of the algebra $W$ 
graded as in Section \ref{sec:weyl}.
\end{lemm}
\begin{demo} We have to check the
$F_0 (W_n)\subset X_n $.
Because $F_0$ preserves the $\star-$product, it is enough to
check that $F_0(W_1)\subset X_1$.
It only means that the (usual) principal symbol of $F(a)$ vanishes
at the origin if $a $ does. It is consequence of Egorov Theorem.

\end{demo}

Summarizing, we have constructed a group morphism
$\ga $ from the group $G$ of elliptic FIO's whose associated
canonical transformation fixes the origin
in the group  ${\rm Aut }(W)$.

\begin{defi}
An automorphism $\Phi $ of $W$ is said to be {\rm real }
($\Phi\in {\rm Aut}_\R (W)$)  if the
mapping $\Phi~ {\rm mod}(\hbar W)$ is real.
\end{defi}
\begin{theo}
The image of the group $G$ by  the homomorphism $\ga $ is
  $ {\rm Aut }_\R (W)$. In particular,
any $\Phi \in {\rm Aut }_\R (W)$ can be ``extended'' to
a semi-classical Fourier Integral
 Operator.
\end{theo}
\begin{demo} The image of $\ga $ is  in the sub-group 
${\rm Aut}_\R (w)$ because the canonical transformation $\chi $
is real.

We have still to prove that the image of $\ga $ is ${\rm Aut }_\R (W)$.
Using the Theorem \ref{theo:main}
and the metaplectic representation, it is enough
to check that the automorphism
${\rm exp }(i{\rm ad}S/\hbar )$ comes from an FIO.
Let $H$ be a symbol  whose Taylor expansion is $S$ (the principal
symbol  of $H$ is a real valued Hamiltonian). 
The OIF $U={\rm exp}\left( i \hat{H}/\hbar \right) $
will do the job.

\end{demo}

\end{document}